# Taming of Modulation Instability by Spatio-Temporal Modulation of the Potential

S. Kumar[1], R. Herrero[1], M. Botey[1], K. Staliunas[1,2]

[1]*Departament de Fisica i Enginyeria Nuclear, Universitat Politècnica de Catalunya (UPC), Colom 11, E-08222 Terrassa, Barcelona, Spain*
[2]*Institució Catalana de Recerca i Estudis Avançats (ICREA), Passeig Lluis Companys 23, E-08010, Barcelona, Spain*

Spontaneous pattern formation in a variety of spatially extended nonlinear system always occurs through a modulation instability: homogeneous state of the system becomes unstable with respect to growing modulation modes. Therefore, the manipulation of the modulation instability is of primary importance in controlling and manipulating the character of spatial patterns initiated by that instability.

We show that the spatio-temporal periodic modulation of the potential of the spatially extended system results in a modification of its pattern forming instability. Depending on the modulation character the instability can be partially suppressed, can change its spectrum (for instance the long wave instability can transform into short wave instability), can split into two, or can be completely eliminated. The latter result is of especial practical interest, as can be used to stabilize the intrinsically unstable system.

The result bears general character, as it is shown here on a universal model of Complex Ginzburg-Landau equations in one and two spatial dimension (and time). The physical mechanism of instability suppression can be applied to a variety of intrinsically unstable dissipative systems, like self-focusing lasers, reaction-diffusion systems, as well as in unstable conservative systems, like attractive Bose Einstein condensates.

Modulation Instability (MI) is at the basis of spontaneous spatial pattern formation in a wide variety of spatially extended nonlinear systems [1-4]. In initial stage of pattern formation, a homogeneous state loses its stability with respect to the modes of spatial modulation. If the growth of unstable modes saturates, stationary and regular patterns develop. Generally, the growth of unstable modes does not saturate, leading via secondary bifurcations to spatio-temporal periodic and chaotic regimes [1-4]. Despite the variety of spatial patterns in nature, the very onset of spatio-temporal dynamics originates from the MI, i.e. from the initial breaking of the maximally symmetric homogeneous state. This is valid both for dissipative nonlinear systems (e.g. chemical, biological, optical systems) and conservative systems (from optical filamentation in Kerr-nonlinear media, instabilities of attractive Bose condensates to presumably, the formation of "rogue waves" [5]). Therefore, manipulating MI would allow for the control of the very onset of unstable pattern dynamics in numerous systems. By manipulation, we mean either decreasing the range of unstable wavenumbers (partial suppression) or completely eliminating the instability. Partial suppression of MI by a purely spatial modulation of the potential has been reported for different systems [6,7]. In some cases this, along with additional stabilization due to finite size of the system (finite trap size) can even inhibit the instability, as e.g. for a Bose Einstein condensate in (stationary) optical lattices [7].

In this letter, we show that a **complete suppression** of MI in infinitely extended systems is possible only by a *spatio-temporal* modulation of the system's potential, satisfying a specific resonant condition.

MI can be universally described by the Complex Ginzburg-Landau Equation (CGLE). The CGLE has been derived systematically as the order parameter equation for many pattern-forming systems close to a Hopf bifurcation, such as the laser [8], chemical systems [9], polariton condensates [10], and others. The CGLE can also be considered as a normal form of the Hopf bifurcation, phenomenologically derived from the assumption of homogeneity of space and time [1-4]; being therefore a universal model. We consider in the main part of the letter one spatial dimensional (1D) CGLE, written in the form:

$$\partial_t A = (1-ic)(1-|A|^2)A + (i+d)\partial_{xx}A \qquad (1)$$

where the threshold parameter and the dispersion coefficient are normalized to unity by scaling time and the space coordinates, respectively. The CGLE contains two free parameters: the coefficient of the cubic (Kerr) nonlinearity $c$, and the diffusion coefficient $d$. In pursuit of clarity, we consider the symmetric case with $d=0$. However, the stabilization can be generally extended to positive (negative) $d$ values, where the CGLE is more (less) stable as compared to the case of $d=0$.

The linear stability analysis is a standard procedure [2, 3], where the stationary, non-zero homogeneous solution of the CGLE, Eq. (1), is subjected to a small perturbation in the form: $A(x,t) = 1 + a(t)\cos(kx)$, $a \ll 1$. The linearization leads to the spectrum of the Lyapunov growth exponents:

$$\lambda_{1,2}(k) = -1 \pm \sqrt{1 + 2ck^2 - k^4} \qquad (2)$$

MI occurs for $c>0$ in Eq. (1), and the range of unstable modes, with real, positive growth exponents (wavenumbers) is: $0 < k^2 < 2c$. A closer inspection of the stability analysis indicates that, for a fixed $c$, the character of MI solely depends on the shape of the spatial dispersion profile of the system. On a zero background, the Laplace operator $i\partial_{xx}^2 A$ in Eq. (1) introduces a parabolic dispersion, $\omega(k) = -k^2$, where $\omega(k)$ represents the eigenfrequency spectrum of the spatial modes, some of which lie within the unstable frequency range, see Fig. 1(a). Assuming that the dispersion of the system can be modified, we now generalize the CGLE, in the form of Eq. (1), for an arbitrary operator of the dispersion $g(\partial_{xx})$:

$$\partial_t A = (1-ic)(1-|A|^2)A + i\hat{\omega}(\partial_{xx})A \qquad (3)$$

where the spatial differential operator $\hat{\omega}(\partial_{xx})$ has a corresponding spectrum, $\omega(k)$. We consider, without loss of generality, that the dispersion is zero for the neutral mode $\omega(0) = 0$. Then, the stability analysis of the generalized CGLE of Eq. (3) provides:

$$\lambda_{1,2}(k) = -1 \pm \sqrt{1 - 2c\omega(k) - \omega^2(k)} \qquad (4)$$

which implies that the instability of modes is entirely determined by the profile of dispersion, $\omega(k)$. The MI condition following directly from Eq. (4) becomes: $-2c < \omega(k) < 0$. This means that the stability of the homogeneous solution of the generalized CGLE depends on the presence of modes with eigenfrequencies $\omega(k)$, within a particular frequency range, as illustrated in Fig. 1(a). Hence, the MI critically depends on the shape of the spatial dispersion curve and could be suppressed or eliminated by modifying the spatial dispersion to prevent the presence of eigenfrequencies within the unstable frequency range (see Fig. 1(b)).

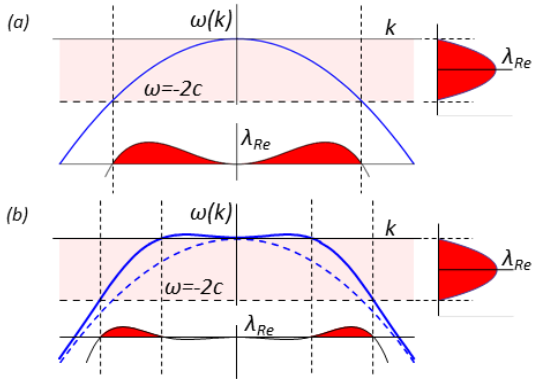

FIG. 1 (color online) (a) MI of the CGLE interpreted in frequency domain: segments of the spatial dispersion curves within $-2c < \omega(k) < 0$, attribute to the MI. A deformation of the spatial dispersion also modifies the MI, transforming it into LW instability in (b). The Lyapunov exponents are calculated by Eq. (4).

The spatial dispersion, $\omega(k)$ can be tailored by introducing a small-scale spatio-temporal periodic modulation of the potential [11]. A similar dispersion tailoring underlies the band-gap solitons [12-15] and subdiffractive solitons [16-18] which are for a defocusing nonlinearity, i.e. for modulationally stable cases.

Thus, combining the idea that the MI depends critically on the spatial dispersion with the fact that a small-scale spatio-temporal periodic modulation of the potential tailors dispersion, we propose that the MI can be manipulated by a proper modulation of the potential, and in the ideal case, can be completely suppressed. The rest of the letter is devoted to substantiate this proposal: starting from a 1D model of CGLE with a modulated potential; we analyze its steady solutions; perform the stability analysis by a modified Floquet approach, and prove the effect by numerical integration of the modulated CGLE in 1D as well in 2D.

For a given physical system, an appropriate modulation of some intrinsic parameter (as in Eq.1) can cause a spatio-temporal modulation of the potential. In general case of CGLE, such a spatio-temporal modulation can be introduced phenomenologically as:

$$\partial_t A = (1-ic)(1-|A|^2)A + i\partial_{xx}A + 4im\cos(qx)\cos(\Omega t)A \qquad (5)$$

where, the modulation is considered on small space and fast time scales, $|q| \gg |k|$ and $|\Omega| \gg |\lambda|$; $k$ and $\lambda$ being the typical wavenumber and growth exponent of instability in unmodulated CGLE. The parameter $m$ is the modulation amplitude.

The steady solution of the modulated CGLE, Eq. (5), according to the Bloch-Floquet theorem, is a periodic function of both space and time, with the periods of the modulation of the potential. Generally, it can be expanded into spatio-temporal harmonics $(n,l)$ of the modulation:

$$A(r,t) = \sum_{n,l} A_{n,l}(t) \cdot \exp(inqx + il\Omega t) \qquad (6)$$

The resonance between harmonics introduce nontrivial effects that can be interpreted as follows: the dispersion curve of the fundamental mode is a parabola centered at ($k=0$, $\omega=0$), while the modulation generates additional harmonics, with corresponding dispersion parabolas shifted horizontally (by $\Delta k = \pm nq$) and vertically (by $\Delta \omega = \pm l\Omega$). To influence the large scale patterns, the modulation parameters $q$ and $\Omega$ must be such that their parabolas cross close to $k=0$ and $\omega=0$. Only three parabolas satisfy this condition in many relevant cases, therefore the expansion (6) can be truncated to the fundamental and the two most significant harmonics shifted by $(\pm q, +\Omega)$. This truncation has been proved useful for various linear [11] and nonlinear systems [16-18]. These three spatio-temporal harmonics [namely $(n,l)$ = (0,0), (-1,-1), (1,-1)] are at, or close to, mutual resonance for $Q = \Omega/q^2 \approx 1$, where we refer to $Q$ as the resonance parameter.

The steady solution (analogous to a Bloch mode), is a locked state of the considered harmonics $A_{0,0}, A_{-1,-1}, A_{1,-1} \neq 0$. In the absence of perturbations it is stable in time for a broad range of parameters; however, close to resonance, $Q \approx 1$, and for sufficiently large $m$ values, it becomes unstable and the harmonics amplitudes develop oscillatory dynamics (black region in Fig. 2a). This attributes to the strong nonlinear coupling between harmonics near resonance.

We perform the linear stability analysis numerically, following the standard Floquet procedure developed for systems homogeneous in space but periodic in time. The evolution of a set of perturbations, introduced to the steady solution, is calculated by integrating over one period of time. As the harmonics of small perturbations at $+k$ and $-k$ are coupled, and their amplitudes are complex, the evolution of 4 independent perturbations have to be numerically calculated for each pair of perturbation modes $a_1(k)$ and $a_2(-k)$. This standard procedure, however, should be modified in view of the spatial modulation of the potential, which causes a linear coupling between the spatial harmonics of the perturbation, $k$, $k\pm q$, $k\pm 2q$... Therefore, such coupled harmonics must be independently perturbed for each $k$, leading to the linear evolution of the ($4n \times 4n$) matrix, $n$ being the number of considered spatial harmonics. Diagonalization of the evolution matrix results in a set of Floquet multipliers whose logarithm gives the average (over a time-period) Lyapunov exponents $\lambda_{Re}(k)$. We typically consider 4 harmonics, checking the convergence for larger number of harmonics.

The result of the stability analysis in parameter space ($Q$, $m$), is summarized in Fig.2(a). The instability is completely suppressed in the central dark-blue island where $\lambda_{Re}(k) \leq 0$ for all $k$ values (Fig.2 (b)). Additionally, we show the regions of partial stabilization with $\lambda_{Re,max} \leq \lambda_0/2$ and $\lambda_{Re,max} \leq \lambda_0/4$, where $\lambda_0$ is the maximal growth exponent of the unmodulated CGLE.

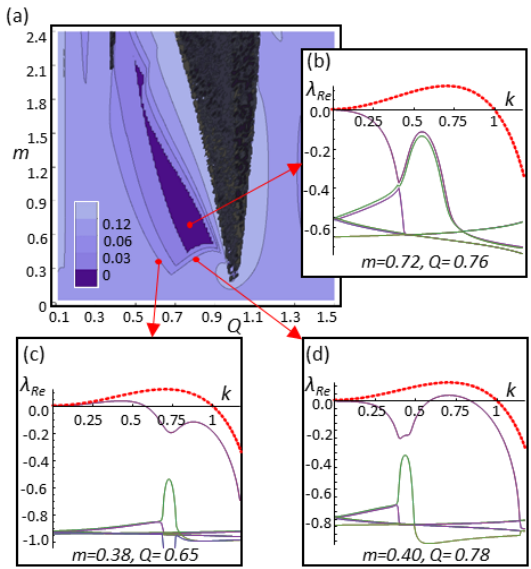

FIG. 2 (color online): (a) Linear stability analysis for $c=0.5$ and $q=4$: non-existence of the steady state solution (black area), complete suppression of MI (dark blue), and partial suppression of MI (light blue). Suppression levels correspond to a factor of 2 and 4 (compared the unmodulated solution). Insets show the spectra of the growth exponents for: complete suppression of MI (b), a typical LW instability (c), and SW instabilities (d). In all cases, the curves with largest exponents are compared to the unmodulated case (red-dashed curve).

The different boundaries of the full stabilization area correspond to different types of remaining weak instabilities: Long Wave (LW) instability (Fig.2 (c)) on the left boundary, and Short Wave (SW) instability (Fig.2 (d)) on the bottom and right boundaries. We note that the stabilization area appears in general for $Q < 1$, approaching $Q \approx 1$ for small $m$ values. This is in accordance with the initial idea that the stabilization occurs close to resonance between the harmonics of plane waves forming the Bloch mode. This basic result holds, qualitatively, for arbitrary $q$ and $\Omega$ (under the limit $|q| \gg |k|$, $|\Omega| \gg |\lambda|$) depending only on the resonance parameter, $Q$.

We confirm the result of the linear stability analysis by numerical integration of the full model (5). We first obtain the stationary Bloch mode (wherever exists, i.e.: outside the black tongue in Fig.2(a)), and then perturb it by a weak random $\delta$-correlated in space perturbation. A sufficiently long integration of (5) (typically 1000 time units) identifies stability/instability of the solution. The calculated stability area perfectly coincides with the region obtained by Floquet analysis in Fig.2 within which all perturbation modes decay, recovering the steady Bloch mode. Outside the stability region, the perturbations grow and the modulated regime sets in. Long-time dynamics is generally complicated (even in the case of unmodulated CGLE, the long-time regime is not completely understood in spite of extensive analysis of last decades [19,20]).

Shown in Fig.3, are the most representative cases of long-time dynamics. For unmodulated CGLE the dynamics is typically chaotic with a spatial spectrum distribution related to the width of the growth exponent spectrum, see Fig.3 (a). The evolution of the modulated CGLE in cases of partial stabilization is typically more "quiet". When LW instability dominates in partially stabilized system, the dynamics is again the typical chaotic one for CGLE, but with correspondingly narrower spatial spectrum, see Fig.3(b). When SW instability remains, we obtain stationary stripe patterns, sometimes interrupted by bursts of random dynamics (see Fig. 3(c)). Generally, the stationary modulated patterns are obtained close to the boundaries of the stabilization balloon.

The stabilization of MI by a modulated spatio-temporal potential can be generalized to more spatial dimensions. We consider the 2D CGLE case, with a modulated potential of rectangular symmetry in space:

$$\partial_t A = (1-ic)(1-|A|^2)A + (i+d)(\partial_{xx}+\partial_{yy})A + \\ 4i(m_x\cos(q_x x)+m_y\cos(q_y y))\cos(\Omega t)A \qquad (7)$$

As in the 1D case, $q_x$ and $q_y$ are the small-scale spatial modulation wavenumbers, and $\Omega$ is the fast temporal modulation frequency ($|q_x|,|q_y| \gg |k|$ and $|\Omega| \gg |\lambda|$). The parameters $m_x$ and $m_y$ represent modulation amplitudes of the potential in the two spatial directions. We consider the simplest most symmetric cases: stripes with $m_x=m$ and $m_y=0$, and squares $m_x=m_y=m$ ($q_x=q_y=q$);

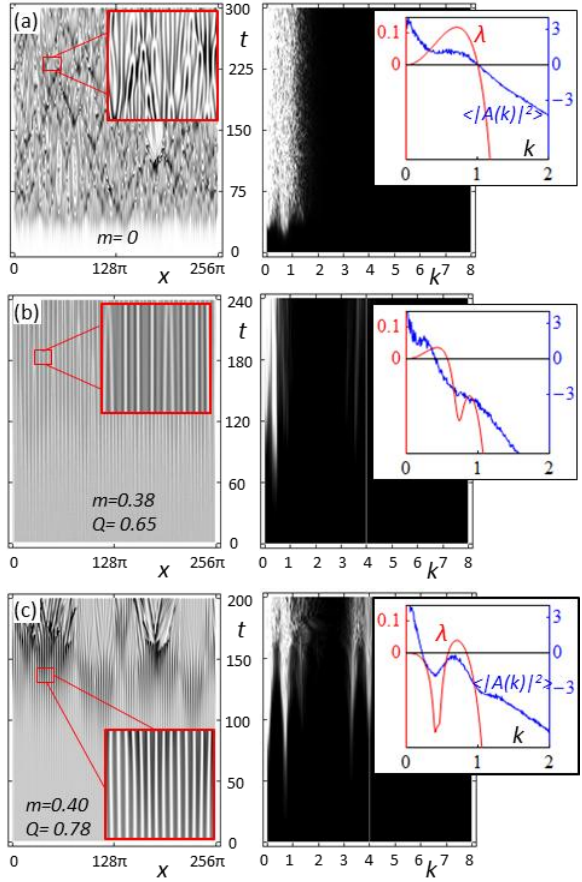

FIG. 3 (color online) Numerically calculated dynamics of the 1D CGLE (a) and partial stabilization for $c=0.5$ and $q=4$ (b, c); with remaining LW instability (b) and SW instability (c). The plots show the intensity (left) and the spatial spectrum (right), plotted against time (in units of $\Omega$). In (c) a stationary pattern (stripe solution) develops. For clarity, small-scale space modulations are filtered out. Insets: upper branch of the spectrum of the Lyapunov exponents (red solid curve; left axis) and long time-averaged $k$-spectrum (blue curve, in logarithmic scale; right axis). The averaging time is 1000.

Simulations by numerical integration of Eq. (6), for different sets of parameters, clearly show the MI elimination and pattern stabilization. We numerically calculate the homogenous solution, perturb it continuously by a $\delta$-correlated perturbation in space and time, and integrate Eq. (6) for a sufficiently long time (typically $t\sim 150$). The field dynamics for the unmodulated case is unstable as it is evident from the spectrum (Fig.4.a). A stripe-potential along horizontal $x$-coordinate partially suppresses the instability in this direction (Fig.4.b), while the square modulated potential can partially (Fig.4.c) or completely (Fig.4.d) suppress the instability, depending on the parameters $m$ and $Q$. Note, that the stabilization area in the 2D case for the squared potential, approximately corresponds to the one obtained in 1D. A detailed exploration of the 2D cases (i.e. for different symmetries of the potential) is outside the scope of this work.

For a visual demonstration of suppression of MI in 2D, a long time ($t\sim 750$) video of the stabilization dynamics is included as supplementary material [†].

Concluding, we propose and demonstrate numerically a mechanism to modify and suppress the MI by a spatio-temporal periodic modulation of potential on small space and fast time scales. We show the phenomenon for a general case, as described by the universal model of the CGLE. We note that in the presence of diffusion ($d\neq 0$) the stabilization effect persists. In fact, a positive diffusion coefficient enlarges the full stabilization balloon (as the instability of unmodulated CGLE with diffusion is weaker than that without diffusion). On the contrary, a negative diffusion coefficient (antidiffusion) reduces the full stabilization balloon. While large antidiffusion prevents complete stabilization, partial stabilization is observed for arbitrary values of the antidiffusion coefficient.

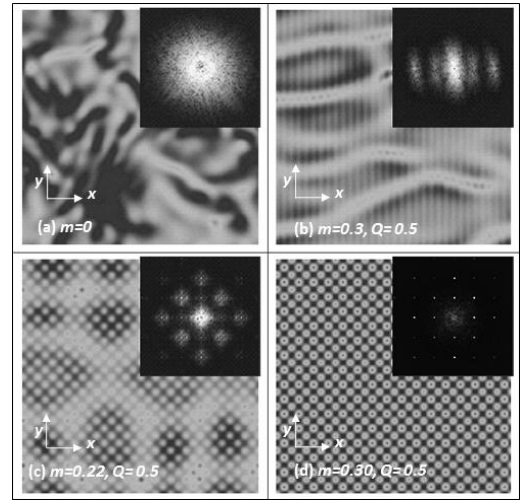

FIG. 4: Numerically calculated patterns in the 2D GCLE showing the field intensity and spatial spectrum (inset) averaged over one time-period after a sufficiently long evolution time ($t\sim 150$). The parameters are $c=0.5$, $d=0.01$ and the size of the integration region is 50x50. Typical chaotic dynamics for the unmodulated 2D CGLE (a). For a spatial modulation only in the horizontal direction (b) the instability is suppressed in that direction but it is evident along the unmodulated coordinate. For a 2D modulation with a square potential, the instability is partially suppressed for parameters lying outside the stabilization area (c) or completely suppressed inside the stabilization area (d). See supplementary material (video) [†] for the full field dynamics.

We note that the modulation of the potential must occur in space and time. Purely spatial modulations, as e.g. in ref. [21] for Bose Einstein condensates in (stationary) optical lattices, do not lead to the proposed stabilization effect. Hence, the presence of a temporal modulation is crucial, as it allows resonant regimes, at $Q \approx 1$, which lie at the basis of the proposed mechanism for the MI suppression.

The proposed effect opens a new possibility towards the stabilization of various spatially distributed nonlinear systems (asymptotically) described by CGLE. Examples in optics are broad aperture lasers and laser-like resonators with Kerr-like focusing nonlinearity, where the MI plays a negative role by destabilizing the emitted radiation. The idea can be also extended to conservative systems, e.g. Bose-Einstein condensates, in order to stabilize the intrinsically unstable attractive condensates. The calculations, however, are to be performed separately for different specific systems.

The work is financially supported by Spanish Ministerio de Educación y Ciencia and European FEDER through project FIS2011-29731-C02-01.

[†] Online link to supplementary video: http://bit.ly/1C0RZdB